\documentclass[twocolumn,prb,amsmath,showpacs]{revtex4}

\usepackage{undertilde}
\usepackage{graphicx}
\usepackage{color}
\usepackage{latexsym}

\begin{document}

\title{Low temperature magnetization
and the excitation spectrum of antiferromagnetic Heisenberg spin
rings}

\author{Larry Engelhardt and Marshall Luban}
\affiliation{Ames Laboratory and Department of Physics and
Astronomy, Iowa State University, Ames, Iowa 50011}

\date{\today}

\begin{abstract}
Accurate results are obtained for the low temperature magnetization
versus magnetic field of Heisenberg spin rings consisting of an even
number $N$ of intrinsic spins $s = 1/2, 1, 3/2, 2, 5/2, 3, 7/2$ with
nearest-neighbor antiferromagnetic (AF) exchange by employing a
numerically exact quantum Monte Carlo method. A straightforward
analysis of this data, in particular the values of the
level-crossing fields, provides accurate results for the lowest
energy eigenvalue $E_{N}(S,s)$ for each value of the total spin
quantum number $S$. In particular, the results are substantially
more accurate than those provided by the rotational band
approximation. For $s \leq 5/2$, data are presented for all even $N
\leq 20$, which are particularly relevant for experiments on finite
magnetic rings. Furthermore, we find that for $s \geq 3/2$ the
dependence of $E_{N}(S,s)$ on $s$ can be described by a scaling
relation, and this relation is shown to hold well for ring sizes up
to $N=80$ for all intrinsic spins in the range $3/2 \leq s \leq
7/2$. Considering ring sizes in the interval $8 \leq N \leq 50$, we
find that the energy gap between the ground state and the first
excited state approaches zero proportional to $1/N^\alpha$, where
$\alpha \approx 0.76$ for $s=3/2$ and $\alpha \approx 0.84$ for
$s=5/2$. Finally, we demonstrate the usefulness of our present
results for $E_{N}(S,s)$ by examining the Fe$_{12}$ ring-type
magnetic molecule, leading to a new, more accurate estimate of the
exchange constant for this system than has been obtained heretofore.
\end{abstract}


\maketitle

\section{Introduction}
\begin{figure}[bt]
\centering
\includegraphics[width=8.5cm]{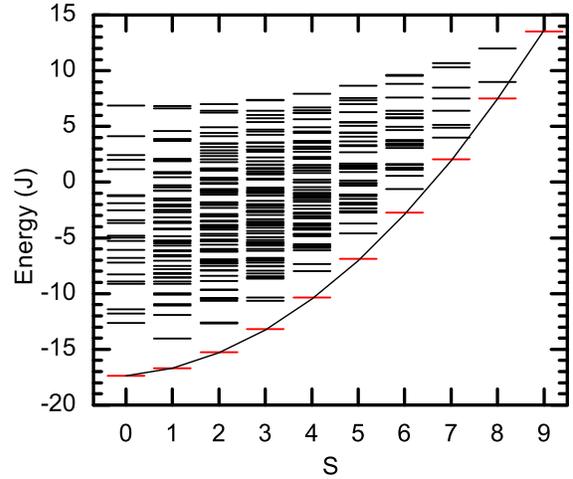}
\caption{(Color online) Complete energy spectrum for $N=6$ and
$s=3/2$ obtained by diagonalizing the Hamiltonain in
Eq.~(\ref{eq:heisenberg}) for $H=0$. The energy levels shown in red
are the subset of minimal energies (SME) for this system as
explained in the text. The solid line is a parabola, given in the
text, that gives an excellent fit to the SME.}

\label{fullSpectrum}
\end{figure}

Since the early 1990s, the field of magnetic molecules has
blossomed, and the number of different species that exist is
increasing rapidly.\cite{nature, science, WPrev1, WPrev2, WPrev3} In
particular, there is a large family of so-called ring-type magnetic
molecules\cite{WPrev1, fe10chem, fe6, fe6fe8, fe18, cr8, fe12chem,
chemRotBand, mn10fe10, cr10, cu10, ni24} that we focus on in the
present work. Within such molecules there are embedded an even
number $N$ of identical paramagnetic ions of intrinsic spin $s$
occupying $N$ equally-spaced sites defining a ring. Each such ion
(``spin'') is coupled to its two nearest neighbors via an AF
exchange interaction, resulting in systems that can
often\cite{fe10chem, fe6,  cr8, fe12chem, chemRotBand, cr10} be well
represented by an isotropic Heisenberg model with a single exchange
energy, $J > 0$, of the form
\begin{equation} \label{eq:heisenberg}
 \utilde{\mathcal{H}} =
 J \sum_{i=1}^{N} \utilde{\vec s}_i \cdot \utilde{\vec s}_{i+1}
 + g \mu_B \vec H \cdot \sum_{i=1}^{N} \utilde{\vec s}_i,
\end{equation}
where the spin operators $\utilde{\vec s}_i$ are given in units of
$\hbar$, $g$ is the spectroscopic splitting factor, and $\mu_B$ is
the Bohr magneton. In the first term of Eq.~(\ref{eq:heisenberg}),
the cyclic character of the system is maintained by requiring that
$\utilde{\vec s}_{N+1} \equiv \utilde{\vec s}_1$. The second term
describes the standard Zeeman effect, where the external field $\vec
H$ is typically defined to be directed along the z-axis. The total
spin operators $\utilde S ^2$ and $\utilde S_z$ then commute with
$\utilde{\mathcal{H}}$, and the eigenstates are described by quantum
numbers $S$ and $M_S$ whose values range from $0$ to $Ns$ and from
$-S$ to $S$, respectively. In Fig.~\ref{fullSpectrum} we display the
zero-field energy spectrum corresponding to
Eq.~(\ref{eq:heisenberg}) for a particular example, $s=3/2$ and
$N=6$, with the subset of minimal energies (SME) shown in red.  The
SME are closely related to what are called level-crossing fields,
quantities which are used to study the SME in great detail
throughout this work for many choices of $s$ and $N$.

In an external magnetic field, the ($2S+1$)-fold degeneracy of each
field-free multiplet is lifted due to a shift, $g \mu_B H M_S$,
originating in the Zeeman term. As the external field is increased
from $H=0$, the ground state will change (among the members of the
zero-field SME) successively from $S=0$, $M_S=0$ to $S=1$, $M_S=-1$,
etc., in integer steps of $S$ and $M_S$ until $S=-M_S=Ns$,
corresponding to saturation of the magnetization. Each of the $Ns$
changes of the ground state quantum numbers is referred to as a
level-crossing, and the field at which the $n$th level-crossing
occurs is denoted in the following by $H_n$.  By determining these
fields, we seek to record the characteristics of the SME as a
function of $s$ and $N$.  This is accomplished using the difference
equation,
\begin{equation} \label{eq:difference}
E_N(S,s) = E_N(S-1,s) + g \mu_B H_n
\end{equation}
for $S=n$, where $n$ extends from 1 to $Ns$.  We elaborate on this
connection between the SME and the $H_n$ in detail in the following
section.

\begin{figure}[t]
\centering
\includegraphics[width=8.5cm]{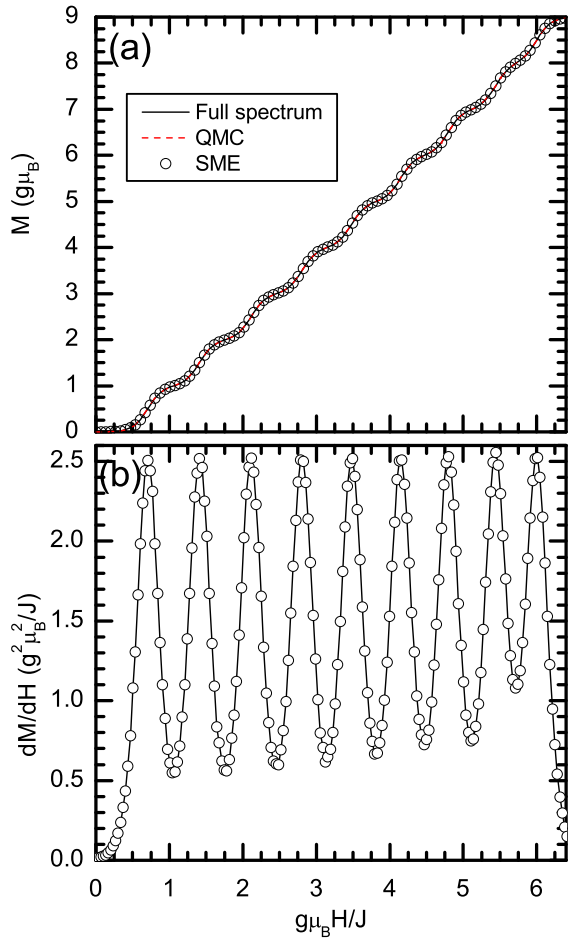}
\caption{(Color online) (a) $M(H)$ and (b) $dM/dH$ for the example
system ($N=6$ and $s=3/2$) at a fixed temperature, $k_B T / J =
0.1$. The data in (b) are obtained by QMC calculations, and as a
guide to the eye the line connects successive data points.}
\label{MvsH}
\end{figure}

In order to appreciate the details of the SME, we first review some
generic features of the spectra, and in particular the SME, that are
already known. It has been noted,\cite{chemRotBand, schnackLuban}
and is clearly evident in Fig.~\ref{fullSpectrum}, that the SME are
accurately approximated by a quadratic dependence on $S$ of the form
$E(S) \propto S(S+1)$, as for a quantum rotor. The solid curve in
Fig.~\ref{fullSpectrum} describes the parabola $E(S) = c J S(S+1) /
(2N) + E_G$, where $c = 4.14$ gives the best fit to this SME which
has a ground state energy $E_G = -17.393 J$. [The reason for the
inclusion of the factor of $2N$ in this equation will become clear
in the next section.] If the SME were strictly parabolic in $S$,
this would give rise to uniformly spaced level-crossing fields.
Although uniform spacing is approximately realized in
Fig.~\ref{MvsH} for our example, we find that the accuracy of such
an approximation deteriorates for larger values of $N$. This is
explored in detail in Sec.~\ref{sec:spectralCoeff}.

Above the SME there exists a large forest of energy levels. Although
many of these levels lie very close to one another, there is a
relatively large energy separation between the SME and the higher
energy levels which has been previously observed.\cite{schnackLuban,
waldmann} Since at a low temperature $T$ only the lowest levels can
be thermally occupied, and all other levels lie well above the SME,
the magnetization as a function of field $M(H)$ consists of a series
of thermally broadened steps that arise at the level-crossing fields
and are determined solely by the SME. [The magnetization is also a
function of $T$, but we will write $M(H)$ for the sake of brevity.]
This step-like property is illustrated in Fig.~\ref{MvsH}, where
$M(H)$ and $dM/dH$, the differential susceptibility, are shown for
the $s=3/2$, $N=6$ example with $k_B T / J = 0.1$. The data in
Fig.~\ref{MvsH}(a) were calculated in three different ways: using
the partition function that includes the complete energy spectrum;
using the quantum Monte Carlo (QMC) method employed in this work;
and using a modified partition function that includes only the
states belonging to the SME. The sharp peaks that appear in
Fig.~\ref{MvsH}(b) were calculated using QMC and the susceptibility
fluctuation formula to give $dM/dH$ directly, not by differentiating
$M(H)$.

The three data sets shown in Fig.~\ref{MvsH}(a) are all identical to
4 significant figures, supporting the assertion that the SME are
sufficient for analyzing low temperature experimental data of this
type. For larger values of $N$, lower temperatures are needed in
order to obtain this degree of agreement, especially in the vicinity
of the saturation field. For this reason, we have carefully checked
that as the temperature is lowered the level-crossing fields have
indeed converged to their limiting, temperature-independent values.
As shown in Sec.~\ref{sec:spectralCoeff}, it is these fields that
are then used to calculate the SME function $E_N(S,s)$.

Despite the very simple appearance of the Hamiltonian of
Eq.~(\ref{eq:heisenberg}), the evaluation of the corresponding
energy eigenvalues and resulting thermodynamic properties frequently
presents a major challenge. The most straightforward way to deal
with this Hamiltonian, and the method that is usually employed when
analyzing magnetic molecules, is to numerically diagonalize the
Hamiltonian matrix.  This yields energy spectra such as that shown
in Fig.~\ref{fullSpectrum}. However, even for relatively small rings
the dimensionality of the Hilbert space, given by $D = (2s+1)^N$, is
so large that the exact diagonalization of the Hamiltonian matrix
becomes totally impractical. For the small ring that has been
considered as an example, $s=3/2$ and $N=6$ gives $D = 4096$. If we
consider a larger ring, for example $s = 5/2$ and $N=12$ which will
be analyzed in Sec.~\ref{sec:fe12example}, we already have $D
\approx 2.2 \times 10^9$, which is well beyond the practical limit
of existing computers.  For $s=5/2$ and $N=20$, $D$ is a staggering
$3.6 \times 10^{15}$.

We can entirely avoid the obstacles confronting matrix
diagonalization by using a QMC method that is not restricted by the
dimensionality of the Hilbert space. We here only focus on low
temperature $M(H)$ and $dM/dH(H)$ which are used to determine the
SME, but other thermodynamic quantities such as the temperature
dependent susceptibility, specific heat, and internal energy are
also readily attainable using this method for all temperatures and
fields, and are in fact computationally much less demanding than the
present low temperature studies.

As seen above, knowledge of the SME enables one to obtain accurate
values of low-temperature $M(H)$ and $dM/dH(H)$ data. To this end,
the SME are calculated in Sec.~\ref{sec:spectralCoeff} for all $s$
from 1/2 to 5/2 and all even $N$ from 4 to 20.  These data are
presented in the form of convenient, dimensionless ``spectral
coefficients'' that will be introduced in that section. The spectral
coefficients are also presented for larger rings, $N=40$, 80, and
larger intrinsic spins, $s=3$ and $s=7/2$. Such large values of $N$
and $s$ have not yet, to our knowledge, been realized in magnetic
molecules, but are useful for studying the approach to the classical
limit ($s \to \infty$).

In Sec.~\ref{sec:gap} the energy gap $\Delta_s(N)$ between the $S=0$
ground state and the lowest $S=1$ state, which can be inferred from
the first level-crossing field, is analyzed in greater detail for
successively larger values of $N$, up to $N=50$, for $s=3/2, 2,
5/2$. This gap is experimentally relevant for NMR and INS
experiments, and is also important for analyzing low temperature,
low-field susceptibility data. Finally, as an illustration of the
usefulness of the present results, in Sec.~\ref{sec:fe12example} we
analyze an existing ring-type magnetic molecule\cite{fe12chem}
composed of 12 Fe$^{3+}$ ions ($s = 5/2$), leading to an improved
estimate for the exchange constant. With the experimental
advancements that are being made both in the synthesis of molecules
and in high field magnetization studies, we anticipate that the use
of the theoretical data presented in this work will complement
future experiments in a much needed way, providing more accurate
estimates of microscopic parameters for future ring-type molecules.

\section{Spectral Coefficients} \label{sec:spectralCoeff}

Since the Hilbert space associated with $ \utilde{\mathcal{H}}$ is
often too large to allow diagonalization of the Hamiltonian matrix,
other theoretical methods must be found. To analyze low-field
susceptibility data, classical spin models and scaled-up data from
smaller systems can sometimes be useful.\cite{fe10chem, fe12chem,
chemRotBand, cr10} However, the level-crossings that are observed in
high-field experiments have no classical analog and cannot be easily
scaled up. For this reason, reliable theoretical data has previously
been lacking, and a main goal of the current work is to remedy this
situation through detailed QMC calculations.

In order to learn about the nature of the SME, we used the
Stochastic Series Expansion method\cite{sse91, sse02} to
simultaneously calculate both $M$ and $dM/dH$ versus $H$ at fixed
temperatures, an example of which was shown in Fig.~\ref{MvsH}. From
these data, we can very accurately infer the level-crossing fields
and thereby reconstruct the SME.  This follows from
Eq.~(\ref{eq:difference}) which gives
\begin{equation} \label{eq:SMEwithH}
E_N(S,s) = g \mu_B \sum_{n=1}^{S}H_n + E_G, \qquad (1 \leq S \leq
Ns),
\end{equation}
where $E_G \equiv E_N(0,s)$ is the ground state energy.  It is
convenient to define the quantities,
\begin{equation} \label{eq:cnDefn}
h_n \equiv \frac{g\mu _{B}H_{n}}{J} = \frac{c_{n}(N,s)n}{N}, \qquad
(n = 1, \ldots, Ns),
\end{equation}
where the dimensionless numbers $c_{n}(N,s)$ will be referred to as
``spectral coefficients''. The energy spectrum of the SME may thus
be written as
\begin{equation} \label{eq:modifiedBand}
E_{N}(S,s) = \frac{J}{N}\sum_{n=1}^{S} n c_{n}(N,s) + E_G, \qquad (1
\leq S \leq Ns). \tag{3$'$}
\end{equation}
Note that if $c_n(N,s)$ were independent of $n$ and given by
$c(N,s)$, Eq.~(\ref{eq:modifiedBand}) would reduce to
\begin{equation} \label{eq:rotband}
E_{N}(S, s) = \frac{c(N,s)J}{N}\frac{S(S+1)}{2} + E_G, \qquad (0
\leq S \leq Ns),
\end{equation}
the so-called ``rotational band'' model that has often been employed
to analyze magnetization data.\cite{fe10chem, fe12mag, fe10nmr,
schnackLuban} Inspecting Eq.~(\ref{eq:cnDefn}), the rotational band
model immediately implies that the level-crossing fields are equally
spaced which, as we will demonstrate in the subsequent subsections,
is hardly the case. Instead, Eq.~(\ref{eq:modifiedBand}), in
conjunction with the spectral coefficients presented in the
following subsection, provides a highly accurate, yet convenient
means of representing $E_N(S,s)$ and thus for analyzing low
temperature magnetization data.

Based on previously known properties of Heisenberg rings, it is easy
to show that $c_{n}(N,s)$ is exactly 4 for a very few special cases.
These are listed here and will be useful in discussing the results
of our calculations in subsequent subsections:

\begin{itemize}

\item[I.] In the case of the $N=4$ ring, $c_{n}(4,s) = 4$,
independent of $n$ and $s$. This is easily derived by describing
this system in terms of two sublattices, each consisting of two
spins. As a result, the SME is given exactly by
$E_{4}(S,s)=JS(S+1)/2 + E_G$.
\item[II.] In the limit of classical spins,\footnotemark $\lim_{s \to \infty} c_{n}(N, s)
= 4$, for all $n$ and $N$.
\item[III.] In all cases $c_{n=Ns}(N,s) = 4$,
independent of $N$ and $s$. This follows from
Eq.~(\ref{eq:modifiedBand}) and the fact that the state with $S =
Ns$ has energy $E_N(Ns,s)=JNs^2$, while the SME energy with $S = Ns
- 1$ is $E_{N}(Ns-1,s)=Js(Ns-4)$.\cite{schmidt}
\end{itemize}

\footnotetext{For rings of classical spins with $N$ even, the SME
can be described by the continuous function $E_N(S_c) = 2J_c S_c^2/N
+ E_G$, given in Eq.~(80) of Ref.~\onlinecite{schmidtLuban}.
Replacing the classical exchange constant $J_c = s(s+1)J$, and the
classical spin $S_c$ by their quantum analogs, we obtain $E_{N}(S,s
\to \infty) = 2JS(S+1)/N + E_G$ from which item II follows.}

Since the spectral coefficients have a value of exactly 4 both in
the limit of very small rings (item I) and in the limit of very
large intrinsic spins (item II), one might expect that the
replacement, $c_n(N,s) = c(N,s) \approx 4$, independent of $n$,
would provide a very good approximation, for example, for Fe$^{3+}$
ions ($s = 5/2$) in small rings ($N \leq 20$). However, as shown in
the following subsections for different choices of $s$ and $N$, the
spectral coefficients do vary significantly with $n$.

\subsection{$s = 1/2$, and $s = 1$} \label{sec:spinOneHalfOne}

Rings of $s=1/2$ spins have been studied using many methods, and a
great deal is known about their spectra. In the 1960s, the lowest
energies, $E_{N=\infty}(S,s=1/2)$, were calculated\cite{griffiths}
in the thermodynamic limit using the Bethe ansatz,\cite{bethe,
betheProof} while numerical diagonalization\cite{bonnerFisher} was
carried out on finite rings. More recently, work has continued for
finite $N$ using methods that include the quantum Monte
Carlo\cite{qmcMvsH}, renormalization group\cite{dmrg, tmrg},
Lanczos\cite{lanczos, waldmann} and conformal field theory
methods.\cite{qft1, qft2}

\begin{figure}[tb]
\centering
\includegraphics[width=8.5cm]{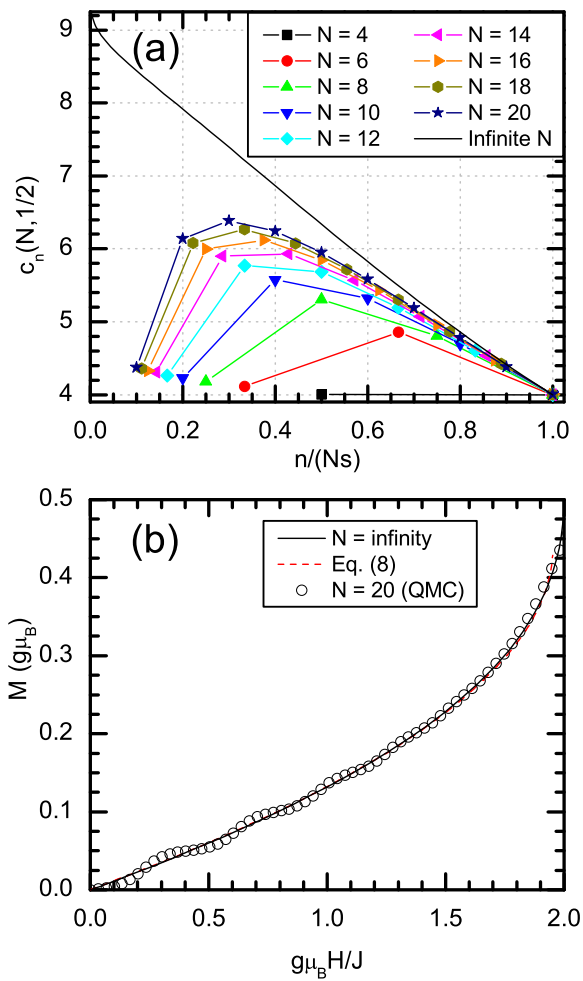}
\caption{(Color online) (a) Spectral coefficients for small $s=1/2$
rings. The solid lines are included to guide the eye, and the
continuous curve corresponds to the $N=\infty$ magnetization data of
Ref.~\onlinecite{griffiths}. (b) $M(H)$ for $N=\infty$ (from
Ref.~\onlinecite{griffiths}), for $N=20$ (QMC) and an approximation
based on (\ref{eq:linear}).} \label{combined0p5}
\end{figure}

The lowest eigenvalues for small $s=1/2$ rings can be easily
obtained from straightforward matrix diagonalization, but are
included here both for completeness and to assess the usefulness of
Eqs.~(\ref{eq:modifiedBand}) and~(\ref{eq:rotband}). The spectral
coefficients that are shown in Fig.~\ref{combined0p5}(a) as a
function of $n/(Ns)$ define the SME for small $s=1/2$ rings.  One
can immediately notice that the $c_n$ vary with $n$, and most are
much larger than 4, implying that a rotational band approximation
provides a relatively poor approximation to these spectra.

Also included in Fig.~\ref{combined0p5}(a) are the spectral
coefficients corresponding to Griffith's original $M(H)$ result for
the infinite $s=1/2$ chain\cite{griffiths} which is shown as a solid
curve in Fig.~\ref{combined0p5}(b). In the thermodynamic limit, the
transformation from magnetization to spectral coefficients can be
accomplished by making the replacement, $n/N \to m_0$, where $m_0$
is the zero temperature magnetization per spin in units of $g
\mu_B$. Eq.~(\ref{eq:cnDefn}) can then be rewritten,
\begin{equation} \label{eq:Mzero}
c_{n}(N=\infty, s) = \frac{h_n}{m_0(h_n)}.
\end{equation}

As can be seen in Fig.~\ref{combined0p5}(a), for $N=\infty$ the
spectral coefficients form a nearly linear function of $n/(Ns)$ over
a very large range of this variable.  Approximating this data as a
linear function,
\begin{equation} \label{eq:linear}
c_{n}(N = \infty ,1/2) \approx \alpha - \beta \frac{n}{Ns},
\end{equation}
and substituting Eq.~(\ref{eq:linear}) into Eq.~(\ref{eq:Mzero}),
again replacing $n/N \to m_0$,
the resulting approximate magnetization is
\begin{equation} \label{eq:sqrt}
m_0 = \frac{\alpha s}{2 \beta} \left(1 - \sqrt{1 - \frac{4 \beta}{s
\alpha ^2}h} \right),
\end{equation}
where $s=1/2$. Fitting the $c_n(\infty,1/2)$ data to the linear
function, we find $\alpha = 8.9$ and $\beta = 5.07$. The
corresponding curve terminates at the point ($h = 1.953$, $m_0 =
0.439$), rather than at (2, 0.5), but otherwise is virtually
indistinguishable in Fig.~\ref{combined0p5}(b) from the exact
magnetization (solid curve). This deviation of the terminus is due
to the fact that the linear approximation of Eq.~(\ref{eq:linear})
does not incorporate a small positive curvature of the cluster
coefficients as a function of $n/(Ns)$ as $n$ approaches $Ns$. Also
included in Fig.~\ref{combined0p5}(b) is $M(H)$ for the $s=1/2$,
$N=20$ ring at a temperature $k_B T/J = 0.05$. This data is nearly
identical to that of the infinite ring, except for the existence of
thermally broadened steps associated with level-crossings.

\begin{figure}[bt]
\centering
\includegraphics[width=8.5cm]{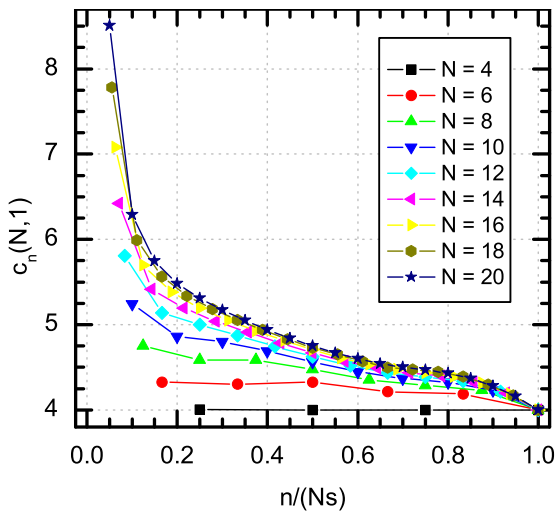}
\caption{(Color online) Spectral coefficients for small rings of
intrinsic spins $s=1$. The solid lines are included to guide the
eye.} \label{1p0coeff}
\end{figure}

Heisenberg rings of $s=1$ spins have received a great deal of
attention since Haldane's prediction\cite{haldane} that a finite gap
separates the ground state from the first excited state in infinite
rings of integer spins $s$.\cite{dmrgGap} In the notation of the
present work, this gap is given by $\Delta_s(N) \equiv E_{N}(1,s) -
E_G = \frac{J}{N}c_{1}(N,s)$, and the values of $c_{1}(N,1)$, seen
as the left-most points in Fig.~\ref{1p0coeff}, are in good
agreement with published values\cite{lanczosGap} of $\Delta_1(N)$.
Values of $\Delta_s(N)$ for all $s$ in the range $1/2 \leq s \leq
5/2$ will be discussed in Sec.~\ref{sec:gap}.

Note however that the data presented here and in the next subsection
include not only the first energy gap [associated with $c_1(N,s)$],
but \emph{all} energy levels that belong to the SME. Studying the
details of of the SME, we find a very rich structure. For instance,
it is evident in Fig.~\ref{1p0coeff} that $c_{n}(N,1)$ decreases
rapidly with increasing $n$, unlike the corresponding data for
$s=1/2$. For $n \geq 0.4 Ns$ the value of $c_{n}(N,1)$ has already
fallen below 5 for $N \leq 20$, whereas for $s=1/2$ this value is
not reached until $n > 0.75 Ns$. In this sense, increasing $s$ from
1/2 to 1 is a significant step on our way toward the classical
limit, stated in item II of the previous section.

\subsection{$s = 3/2$, $s = 2$ and $s = 5/2$} \label{sec:bigSpins}

\begin{figure}[tb]
\centering
\includegraphics[width=8.5cm]{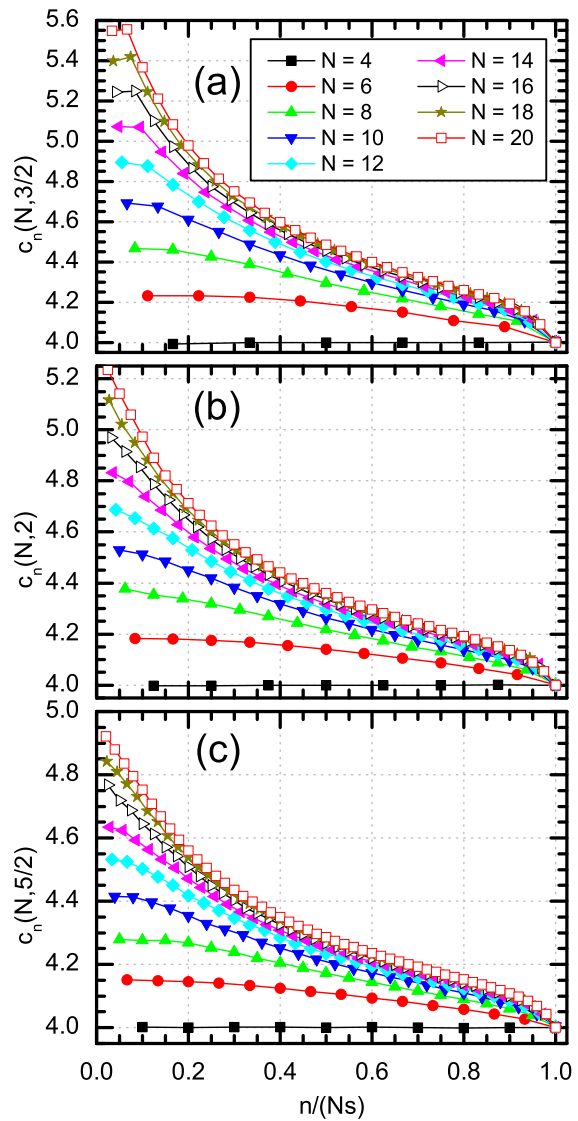}
\caption{(Color online) Spectral coefficients for small rings of
intrinsic spins $s=3/2, 2, 5/2$. The solid lines are included to
guide the eye.} \label{coeff_to_2p5}
\end{figure}

Systems of larger intrinsic spins have also been studied in recent
years,\cite{parkinsonBonner, schnackLuban, waldmann, bigS, schnack}
but with less frequency than $s=1/2$ and $s=1$ systems. Since a
knowledge of the spectral coefficients for $s=3/2$, 2 and 5/2 is
important for a number of molecular rings, these data are presented
in Fig.~\ref{coeff_to_2p5} for all $N \leq 20$. The values of
$c_{1}(N,s)$ that appear in Figs.~\ref{coeff_to_2p5}(a)
and~\ref{coeff_to_2p5}(b) agree with the values of $\Delta_s(N)$
that have been published\cite{schnack} ($N \leq 10$). Again, besides
the first energy gap, the SME exhibit several interesting
characteristics which are reflected in the spectral coefficients.

Of course, the spectral coefficients for $N=4$ are all equal to 4 as
required by item I.  As $N$ increases with fixed $s$ and $n/(Ns)$,
the corresponding spectral coefficients increase from 4
monotonically, resulting in the series of nonintersecting curves
seen in Fig.~\ref{coeff_to_2p5}. This is consistent with Waldmann's
observation\cite{waldmann} that the rotational band model becomes
poorer for larger rings.

\begin{figure}[tb]
\centering
\includegraphics[width=8.5cm]{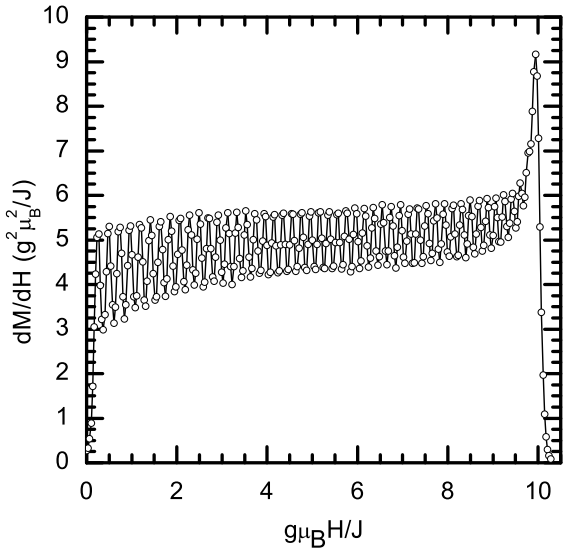}
\caption{Differential susceptibility for a ring of $N=20$ intrinsic
spins $s=5/2$ at a temperature, $k_B T/J = 0.05$. The large peak
immediately before saturation ($g \mu_B H / J=10$) is discussed in
the text.} \label{dMdH}
\end{figure}

Anchored at 4 for $n = Ns$ (item III), and always approaching 4 from
above, the values of the spectral coefficients decrease sharply as
$n$ approaches $Ns$. This ubiquitous drop can be discussed in a
number of contexts. Recalling Eq.~(\ref{eq:cnDefn}), this is clearly
equivalent to a compression of the level-crossing fields as
saturation is approached.  At low temperatures this results in a
large slope of $M(H)$, as can be seen in Fig.~\ref{dMdH} for $N =
20$ and $s = 5/2$. In terms of the energy spectrum, this implies
that the curvature of the SME decreases for large $S$.

Finally, note that as $s$ is increased with fixed $N$ and $n/(Ns)$,
the spectral coefficients descend toward 4 (item II), but only very
slowly. Even for $s=5/2$, most of the spectral coefficients shown in
Fig.~\ref{coeff_to_2p5}(c) are considerably larger than 4,
indicating that one is still far from the classical limit that is
stated in item II. This behavior is explored in the next subsection
with the inclusion of larger values of intrinsic spin.

\subsection{Scaling relation for large $s$} \label{sec:classical}
Thus far we have presented the spectral coefficients that define the
SME as a function of three variables, $s$, $N$, and $n/(Ns)$, and
some general trends have emerged. Now, considering larger values of
$s$ and $N$, we would like to make more quantitative statements
regarding the functional dependence of $c_n(N,s)$ on these
variables. To that end, we have calculated the spectral coefficients
for values of $s$ up to 7/2 and present that data for $3/2 \leq s
\leq 7/2$.

\begin{figure}[tb]
\centering
\includegraphics[width=8.5cm]{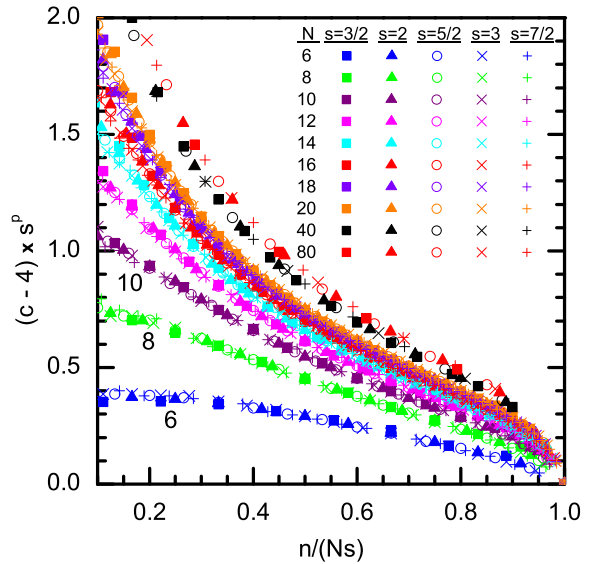}
\caption{(Color online) Spectral coefficients adjusted to
investigate the scaling behavior of Eq.~(\ref{eq:s_scaling}) for all
$s$ in the range $3/2 \leq s \leq 7/2$.  The data shown correspond
to $p=1.05$.} \label{scaled}
\end{figure}

In Fig.~\ref{scaled} we plot the quantity $[c_n(N,s) - 4] \times
s^p$ as a function of $n/(Ns)$ for the choice $p=1.05$. From these
data the $s$ dependence of the spectral coefficients is immediately
evident. For each value of $N$, the data for all $s$ lie on a single
curve, implying that the spectral coefficients scale according to
\begin{equation} \label{eq:s_scaling}
c_n(N,s) = 4 + s^{-p}f(N,n/(Ns)).
\end{equation}
In particular, for $s \to \infty$ Eq.~(\ref{eq:s_scaling}) will be
in accord with item II.  The slow approach to 4 as $s$ is increased
is noteworthy, as even $s = 7/2$ is still far away from the
classical limit. Choosing a slightly different value for the scaling
exponent $p$, such as 1.03 or 1.07, yields visibly inferior results,
so we conclude that $p = 1.05 \pm 0.01$.

A few of the spectral coefficients are also calculated for larger
rings, $N=40$ and $N=80$. The inclusion of this data in
Fig.~\ref{scaled} serves two purposes. First, this data suggests
that $f(N,n/(Ns))$ is indeed converging to a finite limiting curve
in the limit $N \to \infty$, which defines the zero temperature
$M(H)$ of an infinite chain of spins $s$. Secondly, the larger $N$
data strengthens our belief that the scaling relation
(\ref{eq:s_scaling}) is valid for all $N$.

Note that in Fig.~\ref{scaled} data are only included for $n/(Ns) >
0.1$. The data for small $n/(Ns)$ have not been included because the
error in calculating $c_n$ using the QMC method rapidly increases as
$n/(Ns)$ decreases towards zero. The $n=1$ (gap) behavior is
considered in the next section.

\section{Energy Gap} \label{sec:gap}

\begin{figure}[tb]
\centering
\includegraphics[width=8.5cm]{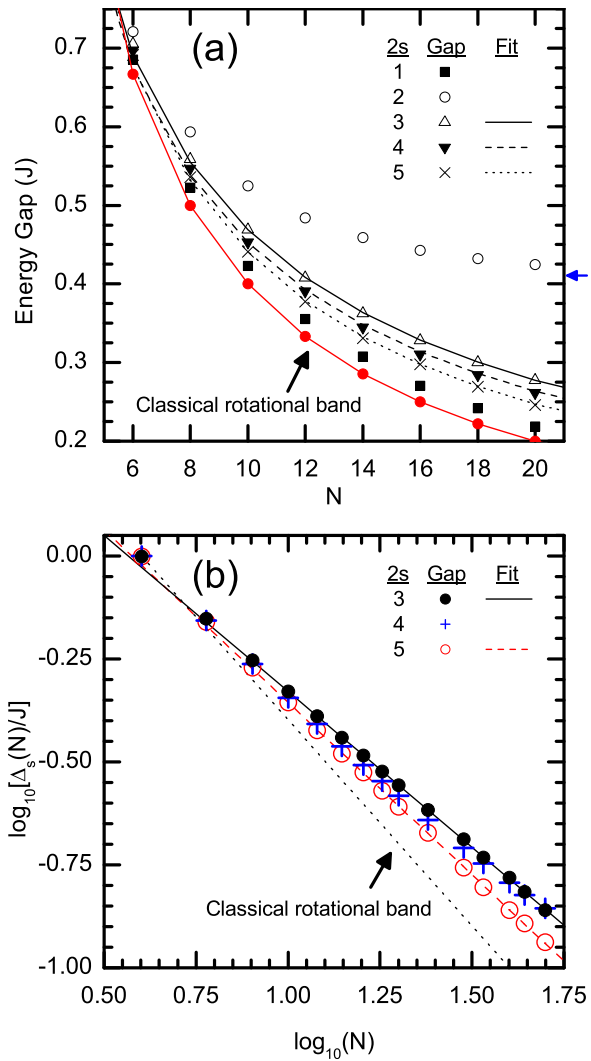}
\caption{(Color online) Energy gap $\Delta _s(N)$ for rings of $N$
spins (a) for all $s \leq 5/2$ (symbols) along with the best fits to
Eq.~(\ref{eq:gapEq}) (curves) as described in the text; and (b) with
$N$ varying from 4 to 50 for $s=3/2$, 2, 5/2 (symbols) along with
the best fits to Eq.~(\ref{eq:gapEq}) (lines) as described in the
text.} \label{gap}
\end{figure}

We now explore the energy gap $\Delta_s(N)$ between the ground state
and the first excited SME level.  Values of this gap are shown in
Fig.~\ref{gap}(a) for rings of $N \leq 20$ spins $s \leq 5/2$. Much
like the behavior of the full SME discussed in the previous section,
this gap systematically approaches the limiting $s = \infty$ form as
$s$ increases from $3/2$, while the $s = 1/2$ and $s = 1$ data
exhibit distinctly different trends.

Specifically, the energy gap for $s = 1/2$ rings is very similar to
the energy gap that would be obtained for a ring with the same value
of $N$ but very large $s$.  This large $s$ limit, indicated in
Fig.~\ref{gap}(a) as the ``classical rotational band'', follows from
item II and is given by $\Delta_{\infty}(N) = 4J/N$. By contrast,
$s=1$ rings have much larger gaps.  Note also that these are already
within 3.5\% of the limiting, $N=\infty$ value even for $N=20$. The
known limiting value,\cite{dmrgGap} $\Delta_1(\infty) \approx 0.4105
J$, is indicated by an arrow on the right side of Fig.~\ref{gap}(a).

Recall that for any $s$, $\Delta_s(4) = J$ from item I. Considering
$N=6$, the classical result $\Delta_{\infty}(6) = 2J/3$ is still a
reasonable approximation to $\Delta_s(6)$, with a relative error of
only a few percent for any $s$. However, as $N$ increases further
this error continues to grow, and with $N=20$ it is nearly 25\% for
$s=5/2$ and nearly 40\% for $s=3/2$.

Although the classical result is not sufficient, we find that the
energy gaps for $s > 1$ are well described by a slightly more
general power law dependence on $N$ of the form
\begin{equation} \label{eq:gapEq}
\Delta_s(N) \sim \Omega N^{-\alpha}.
\end{equation}
The curves in Fig.~\ref{gap}(a) were obtained by choosing: $\Omega =
2.68 J$ and $\alpha = 0.757$ for $s = 3/2$; $\Omega = 2.73 J$ and
$\alpha = 0.781$ for $s = 2$; $\Omega = 3.03 J$ and $\alpha = 0.837$
for $s = 5/2$; while of course $\Omega = 4 J$ and $\alpha = 1$
corresponds to $s = \infty$.  With these choices of $\Omega$ and
$\alpha$, excellent agreement with the QMC data is obtained in the
range $8 \leq N \leq 20$, and it is clear that the classical limit
is indeed being approached with increasing $s$ for both $\Omega$ and
$\alpha$.

For the half odd integer spins, $s=3/2$ and $5/2$, the agreement
with Eq.~(\ref{eq:gapEq}) continues for larger values of $N$. The
same data are shown in Fig.~\ref{gap}(b), now including $N \leq 50$,
and the QMC data agree with the power law formulas to within a
fraction of a percent for all ring sizes in the range $8 \leq N \leq
50$, which is comparable to our uncertainties in $\Delta_s(N)$.  The
values of $\Delta_2(N)$ begin to diverge from the power law
dependence for $N \gtrsim 30$ which is to be expected since they
must eventually converge to a non-zero value. This gap for $s=2$
chains has been previously studied in great detail, and density
matrix renormalization group calculations have yielded a
value\cite{s2haldane} of $\Delta_2(N) = 0.0876J \pm 0.0013J$ in the
limit as $N \to \infty$. One can see in Fig.~\ref{gap}(b) that
$\Delta_2(N)$ is beginning to approach its limiting value, having
become larger than $\Delta_{3/2}(N)$ for $N \geq 50$, but data for
much larger rings would be necessary in order to obtain an accurate
estimate for the limit $N \to \infty$.

The rotational band result, $\Delta_s(N) = 4 J / N$, has been used
in the past\cite{chemRotBand, fe12chem, fe12mag} as an estimate of
$\Delta_s(N)$. Although this provides a reasonable approximation for
$N < 10$, as we have seen it quickly diverges from the correct
result with increasing $N$. As such, it would be prudent to use the
more accurate results presented here when attempting to relate $J$
to the experimentally measured lowest energy gap, e.g., by using
INS, NMR, low-field susceptibility or magnetization data.

\section{An application: F\lowercase{e}$_{12}$} \label{sec:fe12example}

In this section we apply our results to a known magnetic
molecule,\cite{fe12chem} whose analysis has been challenged by a
Hilbert space of dimension $D = 6^{12} \approx 2.2 \times 10^9$. The
molecule is comprised of 12 Fe$^{3+}$ ions ($s = 5/2$), whose
interaction was first investigated\cite{fe12chem} by measuring the
low-field susceptibility $\chi_0(T)$ as a function of temperature
and fitting this data to an approximation of the $s=5/2$, $N=\infty$
chain. The exchange energy thus obtained was $J/k_B=31.9$ K for
$g=2.00$. The field dependent magnetization of the molecule has also
been measured and analyzed, and the first four level-crossing fields
at low temperatures were\cite{fe12mag} $H_1 = 10.1 \pm 0.2 T$, $H_2
= 19.6 \pm 0.2 T$, $H_3 = 29.6 \pm 0.4 T$, $H_4 = 39.1 \pm 0.8 T$.
An analysis of the magnetization was given in
Ref.~\onlinecite{fe12mag} using the classical rotational band
$c(N,s) = 4$, and this yielded the estimate $J/k_B=40.7$ K with
$g=2.02$.  Note that the latter estimate is more than 25\% larger
than the former\cite{fe12chem} estimate. Given the results of
Sec.~\ref{sec:spectralCoeff}, one can expect that the estimate $J =
40.7$ K will be considerably larger than what will result from an
accurate treatment of the Heisenberg model.  This is borne out in
the following.

\begin{figure}[tb]
\centering
\includegraphics[width=8.5cm]{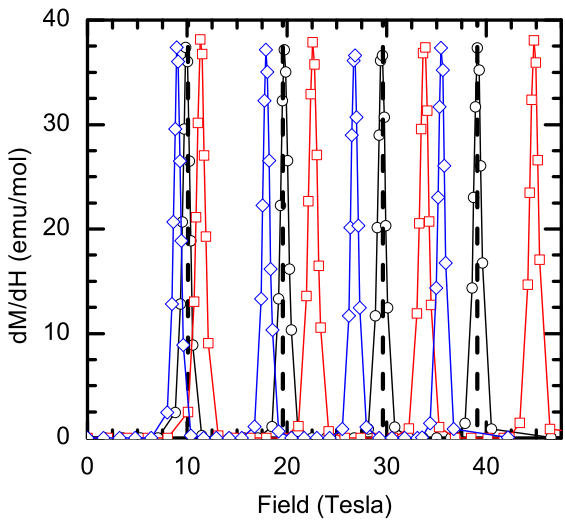}
\caption{(Color online) The four measured level-crossing
fields\cite{fe12mag} (dashed vertical lines) are compared with the
theoretical $dM/dH$ that result from $N=12$, $s=5/2$ Heisenberg
rings with $k_B T / J = 0.01$.  The theoretical data are shown for
the following three choices of $J$ and $g$: $J = 31.9$ K and $g =
2.00$ ({\color{blue} $\diamond$}) from Ref.~\onlinecite{fe12chem};
$J = 40.7$ K and $g = 2.02$ ({\color{red} $\Box$}) from
Ref.~\onlinecite{fe12mag}; $J = 35.2$ K and $g = 2.0$ ($\circ$) are
our best estimates.} \label{fe12chi}
\end{figure}

\begin{figure}[tb]
\centering
\includegraphics[width=8.5cm]{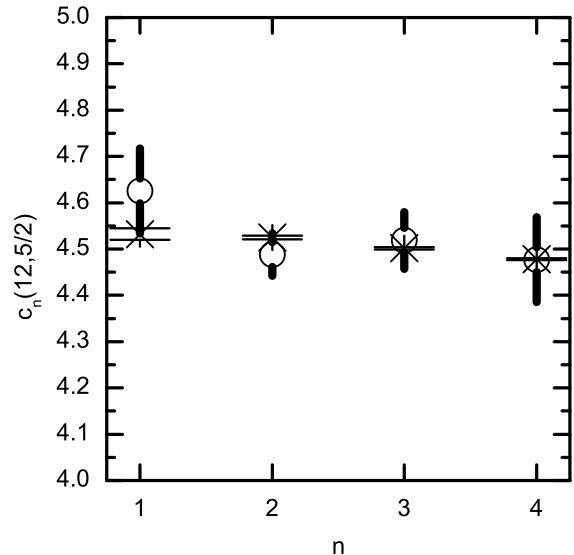}
\caption{The first four spectral coefficients for the $N=12$,
$s=5/2$ ring ($\times$) along with their errors bars, compared with
the results of inserting the first four level-crossing fields and
uncertainties from Ref.~\onlinecite{fe12mag} into
Eq.~(\ref{eq:cnDefn}) with the choice, $J/(k_B g) = 17.6$ K
($\circ$).} \label{Fe12coeff}
\end{figure}

In Fig.~\ref{fe12chi} we compare the four measured level-crossing
fields with our low temperature ($k_B T / J = 0.01$) QMC results. At
this low temperature, each level-crossing of the theoretical
$s=5/2$, $N=12$ Heisenberg ring is clearly indicated by a narrow
peak in $dM/dH$.  Note that the peaks in the QMC data arising from
the parameters $J = 31.9$ K and $g = 2.00$ occur at fields that are
considerably below the experimental level-crossings indicated by the
dashed vertical lines.  On the other hand, the QMC peaks that
correspond to $J = 40.7$ K and $g = 2.02$ are all at fields
significantly greater than the measured values. Particularly at high
fields, these discrepancies become quite pronounced, suggesting that
neither choice of parameters is consistent with the experimental
data. However, we find that the predictions of the Heisenberg model
agree very well with the experimental data if we use $J = 35.2$ K
and $g = 2.0$.  With this choice of parameters, each of the four
theoretical peaks clearly coincides with a measured level-crossing
shown in Fig.~\ref{fe12chi}.

Without using the $dM/dH$ level-crossing field data directly, one
can easily arrive at the same estimate based on the spectral
coefficients of Sec.~\ref{sec:spectralCoeff}. Recalling
Eq.~(\ref{eq:cnDefn}), the ratio of $J$ to $g$ is given by $J/g = N
\mu_B H_n/[n c_n(N,s)]$. An estimate of this ratio for a given
molecule is then obtained by simply inserting a measured value of
$H_n$ and the corresponding $c_n(N,s)$ from Fig.~\ref{coeff_to_2p5}.

Alternatively, from the measured $H_n$ we can construct an
experimental analog of the spectral coefficients by fixing the ratio
$J/g$ in Eq.~(\ref{eq:cnDefn}). In Fig.~\ref{Fe12coeff} we display
those results for the four measured $H_n$ (and their uncertainties).
These data are in good agreement with the spectral coefficients if
we choose the ratio $J/(k_B g) = 17.6$ K, consistent with our
previously stated estimate.

A small decrease with increasing $n$ is observable in the spectral
coefficients derived from the experimental values of the
level-crossing fields. This is expected from the data presented in
Sec.~\ref{sec:spectralCoeff}, but more level-crossings and/or
smaller experimental error bars are needed in order to clarify this
point. These data are also useful for getting a sense of the typical
errors in the spectral coefficients that were presented in
Sec.~\ref{sec:spectralCoeff}.  As shown in Fig.~\ref{Fe12coeff}, the
error bars of the QMC data decrease very rapidly with increasing $n$
and are in fact not visible in Figs.~\ref{coeff_to_2p5}
and~\ref{scaled}.

Our conclusion is that the existing data for the Fe$_{12}$ molecule
is best fit by the choice $g =2.0$, $J=35.2$ K. This value of $J$ is
13.5\% smaller than the value that resulted\cite{fe12mag} from
analyzing the experimental level-crossing fields using c(N,s) = 4.
This reflects the fact that the spectral coefficients, although not
constant, exceed 4 by approximately 13\%. A similar analysis would
be equally straightforward for any other rings whose spectral
coefficients are shown in Sec.~\ref{sec:spectralCoeff}.

\section{Summary} \label{sec:summary}

In this article we have utilized a quantum Monte Carlo (QMC)
method\cite{sse91, sse02} to calculate detailed properties of the
general quantum Heisenberg ring.  This system consists of an even
number $N$ of equally-spaced spins $s$ mounted on a ring, where the
spins interact via nearest-neighbor antiferromagnetic isotropic
exchange, with a single exchange constant $J$.  As this system does
not exhibit magnetic frustration it was possible to calculate
thermodynamic quantities down to very low temperatures.  In this
work our primary focus has been on the accurate determination of the
level-crossing fields, which in turn directly provide the lowest
energy eigenvalue $E_N(S,s)$ for each value of the total spin
quantum number $S$.  By introducing the notation of spectral
coefficients [see Eq.~(\ref{eq:cnDefn})], denoted by $c_n(N,s)$, we
obtained an especially convenient representation of $E_N(S,s)$,
given by Eq.~(\ref{eq:modifiedBand}). As the QMC method operates
without diagonalizing the Hamiltonian matrix, we were able to obtain
results for spins $s=1/2$, 1, 3/2, 2, 5/2, 3, 7/2, focusing mostly
on $N \leq 20$ as these are experimentally relevant, although $N
\leq 80$ were also considered. Among our principal results, we have
found that the set of level-crossing fields are not uniformly
spaced, and thus the spectral coefficients $c_n(N,s)$ vary
significantly with $n$. Equivalently, $E_N(S,s)$ departs from the
strictly quadratic dependence on $S$, referred to as the rotational
band approximation\cite{schnackLuban} (equivalently, the
Land$\acute{e}$ interval rule). These deviations from uniform
spacing are fairly small for $N \leq 8$, however, they become
increasingly severe with increasing $N$.  Similarly, the ground
state energy gap, which may be written as $\Delta_s(N) = E_N(1,s) -
E_N(0,s) = c_1(N,s)J/N$, varies significantly with $N$ and $s$.  For
$s=1$, 2 we find that, consistent with the Haldane
result,\cite{haldane} $\Delta_s(N)$ is indeed converging to a
non-zero limiting gap for large $N$, and in good agreement with
estimates in the literature\cite{dmrgGap,s2haldane} for these two
choices of $s$.  By contrast, for $s=3/2$, 5/2 we find that
$\Delta_s(N)$ appears to decrease to zero for large $N$ according to
a power law, [see Eq.~(\ref{eq:gapEq})], where the exponent is given
by $\alpha \approx 0.76$ for $s = 3/2$ and $\alpha \approx 0.84$ for
$s = 5/2$. The increase with $\alpha$ towards unity with increasing
$s$ is consistent with the known rigorous result,\cite{schmidtLuban}
$\Delta_{\infty}(N) = 4J/N$, for the classical Heisenberg ring,
which may be pictured as the quantum Heisenberg ring in the limit $s
\to \infty$.  We also find that the departure of the general
spectral coefficient $c_n(N,s)$ from the classical result
$c_n(N,\infty)=4$ is characterized by power law behavior [see
Eq.~(\ref{eq:s_scaling})]. Finally, we have illustrated the
practical utility of our present results for the level-crossing
fields and $E_N(S,s)$ by considering the ring-type magnetic molecule
known\cite{fe12chem} as Fe$_{12}$. In particular, our analysis of
the existing\cite{fe12mag} experimental data for level-crossing
fields shows that this system can be very well described by the
nearest-neighbor Heisenberg model with antiferromagnetic exchange,
and we are able to provide a new and improved estimate of the
exchange constant. Although only rings with even $N$ have been
considered in this work, we suspect that similar scaling relations
may hold for other structures as well.

\begin{acknowledgments}
The authors would like to thank H.-J.\ Schmidt and J.\ Schnack for
useful comments.  Ames Laboratory is operated for the United States
Department of Energy by Iowa State University under contract No.\
W-7405-Eng-82.
\end{acknowledgments}

\bibliography{engelhardt_prb}

\end{document}